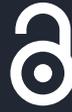

# In Which Sense Can We Say That First-Class Constraints Generate Gauge Transformations?

**ÁLVARO MOZOTA FRAUCA**

## ABSTRACT

In this paper, I consider a recent controversy about whether first-class constraints generate gauge transformations in the case of electromagnetism. I argue that there is a notion of gauge transformation, the extended notion, which is different from the original gauge transformation of electromagnetism, but at the same time not trivial, which allows the making of that claim. I further argue that one can expect that this claim can be extended to more general theories, and that Dirac's conjecture may be true for some physically reasonable theories and only in this sense of gauge transformation. Finally, I argue that the extended notion of gauge transformation seems unnatural from the point of view of classical theories, but that it nicely fits with the way quantum versions of gauge theories are constructed.

**CORRESPONDING AUTHOR:**
**Álvaro Mozota Frauca**

Department of Physics, Universitat de Girona, Carrer de la Universitat de Girona 1, 17003 Girona, Spain

alvaro.mozota@udg.edu

# 1 INTRODUCTION

In the literature dealing with the constrained formalism used for the Hamiltonian treatment of gauge theories, there is a widely accepted claim that first-class constraints generate gauge transformations.[1] This claim is referred to as the Dirac conjecture or Dirac hypothesis,[2] and it has been challenged in the literature. One criticism of this claim was formulated by Brian Pitts (2014),[3] who showed that, in the case of electromagnetism, the electric field as defined in the standard way, that is, $\dot{\vec{A}} - \vec{\nabla} A_0$, doesn't remain invariant under a transformation generated by the secondary constraint of the theory. This criticism has recently been replied to by Oliver Pooley and David Wallace (2022), who showed how even if $\dot{\vec{A}} - \vec{\nabla} A_0$ doesn't remain invariant, the empirical content of electromagnetism can be preserved by correctly identifying the electric field. To this, Pitts has acknowledged (Pitts 2022) that the empirical content can be preserved in the way proposed by Pooley and Wallace, but he argues that this doesn't correspond to a gauge transformation, just to a de-Ockhamization, i.e., a trivial transformation in which one substitutes a quantity with the sum of two other quantities. Furthermore, he argues that, if one accepts the claim by Pooley and Wallace, one also has to accept the claim that (at least some) second-class constraints generate gauge transformations, contrary to the "orthodoxy" Pooley and Wallace aimed to defend.

In this paper, I will analyze this controversy and I will note that much of the argument depends on what counts as a gauge symmetry. In this sense, I will distinguish between three possible notions of gauge symmetry. The first one is to consider gauge transformations to be any transformation that affects the mathematical structures of a theory but leaves its empirical content intact. This is the notion of gauge transformation, which Pitts argued is too broad and generous, as it would include all the artificial de-Ockhamizations that have nothing to do with what physicists refer to when they speak of gauge theories. This notion of gauge transformation allows claiming that first-class constraints generate gauge transformations, but it also entails that second-class constraints also do, contrary to the orthodoxy.

The second notion of gauge transformation is the more restrictive one that Pitts seems to have in mind. This would correspond to the "genuine" gauge transformations, which are the symmetry transformations of theories like electromagnetism or Yang-Mills theories. These theories are formulated not in terms of the directly observable and physical quantities but in terms of some more convenient quantities. For instance, in the case of electromagnetism, the introduction of the 4-potential allows writing Maxwell equations in the form of second-order differential equations or defining the potential energy of a particle in an electromagnetic field. It also allows expressing the theory by making use of the Lagrangian formalism, in which the original gauge symmetry appears as a symmetry of the action. When expressed in the Hamiltonian formalism, this symmetry transformation is generated by the gauge generator, which, as Pitts argued, is a tuned sum of constraints. In this sense, for theories like electromagnetism, we can start with an original notion of gauge transformation, even before expressing the theory in Lagrangian or Hamiltonian form.[4] In the Lagrangian formalism this symmetry is represented by a symmetry of the action, and in the Hamiltonian formalism

---

the symmetry transformations are generated by the generator. For theories like Yang-Mills theories, it is more common that the theories are formulated directly in the Lagrangian formalism, but even here a good case can be made that the "genuine" gauge transformations are symmetry transformations of the action, and hence, that they are generated by the gauge generator. First-class constraints do not generate transformations that are symmetries of the original Lagrangian action, and therefore they do not generate gauge transformations in this restricted sense, as both parts in this debate acknowledge.

Having these two notions of gauge transformation in mind, Pitts (2022) argues that the claim that first-class constraints generate gauge transformations is either trivial (if we take the first notion) or false (if we take the second). In either case, the orthodoxy would be in trouble, as in the first case it would have to admit that second-class constraints generate gauge transformations, and in the second that first-class constraints do not. For this reason, I think that the position Pooley and Wallace ought to take is that first-class constraints generate gauge transformations in a third, intermediate sense, which is at the same time not the original Lagrangian gauge transformation but not a completely trivial de-Ockhamization. It is by adopting this notion of gauge transformation that one can save the orthodoxy and claim that first-class constraints in electromagnetism generate gauge transformations, while second-class constraints in theories like Proca electromagnetism do not.

This third notion of gauge transformations shares with trivial de-Ockhamizations that there appear new compensating fields and variables in the formalism, but contrary to the case of de-Ockhamizations, one can preserve the empirical content with no need to pay much attention to them, or to introduce redefinitions. In this sense, the de-Ockhamizations in this sort of transformation would only affect accessory variables and not the physically relevant and empirically observable ones.

Part of the controversy between Pitts and Pooley and Wallace lies precisely on this point. While Pitts sees the issue from a Lagrangian-first perspective, which takes configuration variables ($A_\mu$ in the case of electromagnetism) as physical and momentum variables ($\vec{\pi}$) as accessory, Pooley and Wallace take a Hamiltonian-first perspective in which they adopt the opposite view, namely, that some momentum variables ($\vec{\pi}$) are physical while some configuration variables ($A_0$) are just accessory. For this reason, it is important to notice that adopting Pooley and Wallace's view breaks with the more natural understanding of classical theories and the way they are represented using the Lagrangian and Hamiltonian formalisms. However, despite this, it is a position one can hold in a consistent way to save the orthodoxy.

The next natural question to ask is which of the three notions of gauge symmetry is best or physically more relevant. From the point of view of classical physics, this seems a strange question, as the physical content of a classical gauge theory lies in its empirical content. For instance, it is Maxwell's equations for the electromagnetic field and the Lorentz force equation, which exhaust all the physical content of electromagnetism. In this sense, classical gauge transformations, in any of the three senses discussed above, do not seem interesting from the point of view of physics but just from the point of view of how we represent physics. From this perspective, one may be tempted to keep gauge transformations at a minimum and keep just the original Lagrangian ones and preserve the Lagrangian-Hamiltonian equivalence. When we move to quantum physics, it may be the case that gauge structures play a more non-trivial role, and that the issue becomes more relevant, but notice that neither Pitts nor Pooley and Wallace discuss the physical relevance of adopting one view of gauge or another.

In this article, I will note there seems to be a reason for adopting the middle-way view of gauge transformations: quantization. When following canonical quantization procedures, one imposes that physical states satisfy the first-class constraints in a way that is equivalent to imposing that they are invariant under transformations generated by them. It is, therefore, very natural to read these transformations as the quantum version of a gauge transformation, and hence this fits very naturally with the middle-way view of gauge transformations in which first-class constraints generate gauge transformations.

To insist, in this article I will argue that there are (at least) three possible notions of gauge transformation:

1.  **Trivial gauge transformations:** Any transformation that preserves the empirical content of the theory, even if one needs to introduce compensating fields and redefine what counts as physical. This includes de-Ockhamizations. In the Hamiltonian formalism, they are generated by any phase space function.

2.  **Genuine or original gauge transformations:** Gauge transformations associated with symmetries of the original, non-de-Ockhamized Lagrangian action or that can be defined even before making use of the Lagrangian formalism. In the Hamiltonian formalism they are generated by the gauge generator.

3.  **Extended gauge transformation:** A transformation that preserves the empirical content of the theory, with no need to redefine physical fields, even if one needs to introduce compensating fields and redefine accessory fields. This notion implies a departure from the original Lagrangian action and symmetries and adopting a view in which conjugate momenta can be physical and configuration variables accessory. In the Hamiltonian formalism, they are generated by first-class constraints.

I will argue that, from the point of view of classical physics, the second sense is more natural, while the third sense is consistent and fits nicely with canonical quantization procedures. I will illustrate this with a discussion of some relevant examples.

Let me clarify that, besides this classification of gauge transformations, there is a further distinction that appears in the literature and that is conceptually independent of most of what is discussed in this article. This is the distinction between global gauge transformation, a transformation that transforms solutions of the equations of motion into solutions of the equations of motion, and local or instantaneous gauge transformation, which transforms the fields at an instant of time or at a spacetime point. For most theories, both ways of thinking about gauge transformations are equally fine, but for reparametrization invariant theories, such as general relativity, it is only from the global point of view that one can speak about gauge transformations.[5] For the examples discussed in this article, we won't have to worry about this distinction, and one can think about gauge transformations either way.

Finally, let me make a methodological remark. While, in the literature, one sometimes finds the claim that it is the gauge structure of a theory that defines its physical content, I will be taking the opposite perspective. That is, I will be discussing examples with clear physical content and then I will be making claims about the different theoretical representations of this physical content and the symmetries of these representations. For instance, I will be discussing the dynamics of free Newtonian particles, which will uncontroversially consist of particles moving with uniform velocities $q(t)$, and then I will be considering different symmetry transformations of possible representations of these dynamics. In the case of the debate about electromagnetism, all the parts seem to agree in that the empirical content of the theory is an electromagnetic field that satisfies Maxwell's equations and which "pushes" electrically charged particles, and then they disagree about what is to be considered gauge.

---

5    See the arguments for this in Maudlin (2002), Mozota Frauca (2023), and Pitts (2017; 2018).

The structure of the article is the following: I will first discuss in more detail paradigmatic examples of genuine gauge transformations and de-Ockhamizations in section 2, but I will introduce some more subtle examples that can be argued to be halfway between the two. These examples can be built by means of a "momentum de-Ockhamization," i.e., by the introduction of an unphysical degree of freedom with the sole aim of compensating for the changes that the putative gauge transformation is introducing in the momentum variables used for describing the system. Despite being built artificially and by the introduction of unphysical degrees of freedom, I will argue that they are closer to "genuine" gauge transformations than pure de-Ockhamizations, as the de-Ockhamization affects the momentum variable that can be considered to be accessory and not the configuration one, which can be considered physical. In this sense, we will find a parallelism with the case of electromagnetism: if de-Ockhamizations affect (arguably) accessory variables, like $p$ in the example I will discuss or $A_0$ in the case of electromagnetism, while leaving intact the (arguably) physical ones, one can argue that the transformation is not of the trivial kind, even if it is not a symmetry of the original, minimal Lagrangian.

Then, I will move to analyze the cases of Maxwell electromagnetism and Proca electromagnetism in section 3. I will follow Pitts in arguing that second-class constraints[6] generate gauge transformations in Proca electromagnetism if one takes a trivial definition of gauge. That is, only if we perform some redefinitions and reidentifications are we able to preserve the empirical content of Proca's theory. I will agree with Pitts in that this is bad news for the orthodoxy: If one relaxes too much the standards for what is considered a gauge transformation, any transformation can count as a gauge transformation, and the orthodoxy wants to maintain that first-class constraints generate gauge transformations, while second-class do not.

Next, I will analyze the case of standard electromagnetism in the extended formalism, and I will argue that a consistent notion of gauge transformation is available, which allows keeping the orthodoxy and claiming that first-class constraints generate gauge transformations in electromagnetism. The price to pay is to lower the standards that Pitts holds, but not as much as to let any transformation be considered a gauge transformation. This notion of gauge still has the consequence that second-class constraints do not generate gauge transformations, and that de-Ockhamizations aren't considered gauge transformations, either. The discussion will illustrate how this notion of gauge transformation depends crucially on what we take to be physical and accessory variables, as it will define gauge transformations to be transformations affecting and even de-Ockhamizing accessory variables, but not the physical ones.

In section 4, I move away from the particular case of electromagnetism and the examples discussed in this article and argue for three general claims. First, I give a more detailed argument supporting the claim that any phase space function, in any theory (gauge or not), can be said to generate a trivial gauge transformation in the form of a de-Ockhamization. Second, I argue that, when the theory is a first-class system and the phase space function or functions chosen are constraints, this procedure leads to an extended formalism in which the symmetry transformations are generated by the first-class constraints. Third, I further argue that this symmetry transformation is plausibly a gauge transformation in the extended sense once some restrictions are in place. In this sense, I argue that it seems possible that Dirac's conjecture can be kept for most physical theories and for the extended notion of gauge, although I do not give rigorous proof for it.

---

6    To be clear, while Pitts argued that it is just the primary constraint of the theory that generates a trivial gauge transformation, I will argue the secondary constraint also generates a trivial gauge transformation, and that indeed any phase space function can be used to generate a trivial gauge transformation.

Finally, in section 5 I will analyze how this affects the quantization of theories. I will argue that Dirac's quantization procedure nicely fits with the extended Hamiltonian formalism with the assumption that observables are invariant under transformations generated by the constraints, given that the way constraints are imposed automatically implies invariance under the quantum counterpart of such transformations. In this sense, even if the extended sense of gauge transformations was a bit unnatural from the classical perspective, it is the sense that fits better with the way quantum theories are built. If someone wanted to stay closer to the original, "genuine" gauge transformations in quantizing the theory, one should devise a different quantization procedure in which constraint imposition does not lead to the extended Hamiltonian notion of gauge symmetry.

# 2 GAUGE, DE-OCKHAMIZATIONS, AND OTHER TRANSFORMATIONS

I will start by giving a series of examples of different transformations that can be given the name of "gauge transformations," even if in different senses. That is, all of them can be interpreted as leaving intact the empirical content described by the models they are applied to, but one may have different intuitions about whether they really are gauge transformations or not.

## 2.1 GENUINE GAUGE SYMMETRY: ELECTROMAGNETISM

The paradigmatic case of genuine gauge transformation is the (original) gauge symmetry of Maxwell electromagnetism, and as it is the central example in the controversy, it will be useful to introduce it now. The physical content of electromagnetism can be summarized in the four Maxwell equations and the Lorentz force equation, which describe the evolution of the electric and magnetic fields, $\vec{E}$ and $\vec{B}$, in the presence of charged matter and the effect of these fields on the dynamics of such matter. However, it is convenient to introduce the 4-vector field $A_\mu$, which relates to $\vec{E}$ and $\vec{B}$ by means of:

$$\vec{E} = \dot{\vec{A}} - \vec{\nabla} A_0 \tag{1}$$

$$\vec{B} = \vec{\nabla} \times \vec{A}, \tag{2}$$

where $\vec{A}$ represents the spatial components of $A_\mu$. This introduces a redundancy, as for each pair $\vec{E}, \vec{B}$ there exists a whole family of 4-potentials that give rise to the same $\vec{E}, \vec{B}$. The gauge transformation of electromagnetism is any transformation that maps from one $A_\mu$ to another in the family. In particular, they can be expressed in the form:

$$A_\mu \rightarrow A_\mu + \partial_\mu \epsilon, \tag{3}$$

where $\epsilon$ is a function of the spacetime coordinates. The introduction of $A_\mu$ allows for expressing the theory in the Lagrangian formalism:

$$S_{em}[A_\mu] = \int dt d^3 x \left\{ \frac{1}{2} (\dot{\vec{A}} - \vec{\nabla} A_0)^2 - \frac{1}{2} (\vec{\nabla} \times \vec{A})^2 - (A_0 \rho + \vec{j} \cdot \vec{A}) \right\}, \tag{4}$$

where $\rho$ and $\vec{j}$ describe the charge density and current. The gauge symmetry of the theory is reflected in the action, as it is invariant under a transformation of the form 3, provided that $\rho$ and $\vec{j}$ satisfy the continuity equation $\dot{\rho} + \vec{\nabla} \cdot \vec{j} = 0$. This feature implies that the Lagrangian is singular[7] and that one needs to use the constrained formalism if one wants to express electromagnetism in the Hamiltonian language.

In particular, this system has a primary constraint $C_1 = \pi^0 = 0$,[8] where $\pi^0$ is the momentum conjugate to $A_0$ and the dynamics of the system can be defined[9] by the total Hamiltonian density:

$$H_T(\vec{A}, \vec{\pi}, A_0, \pi^0) = H_c(\vec{A}, \vec{\pi}, A_0) + \lambda \pi^0 = \frac{1}{2}(\vec{\pi}^2 + \vec{B}^2) - A_0(\vec{\nabla} \cdot \vec{\pi}) + (A_0 \rho + \vec{j} \cdot \vec{A}) + \lambda \pi^0. \quad (5)$$

Imposing consistency of the primary constraint, i.e., $\dot{\pi}^0 = 0$ leads to the secondary constraint $C_2 = \vec{\nabla} \cdot \vec{\pi} - \rho = 0$. Imposing consistency of the secondary constraint, i.e., $\frac{d}{dt}(\vec{\nabla} \cdot \vec{\pi}) = 0$, doesn't lead to any new constraints or to any condition on $\lambda$. This is because the constraints are first-class, i.e., $\{C_1, C_2\} \approx 0$.[10]

The dynamics defined by the Hamilton equations, together with the two constraint equations, are equivalent to the dynamics defined by the Euler-Lagrange equations for the action 4 together with the equation $\lambda = \dot{A}_0$. The Euler-Lagrange equations, as well as the original Maxwell equations, do not imply any dynamical equation for $A_0$, and this is explicit in the total Hamiltonian formalism where $A_0$ is allowed to vary according to the arbitrary function $\lambda$. In this sense, the original gauge symmetry is explicit in the total Hamiltonian equations.

One can find the Hamiltonian version of the gauge transformation of the theory by finding the transformation which is a symmetry of the total action:

$$S_{em}[\vec{A}, \vec{\pi}, A_0, \pi^0] = \int dt d^3x \left( \dot{\vec{A}} \cdot \vec{\pi} + \dot{A}_0 \pi^0 - H_c - \lambda \pi^0 \right). \quad (6)$$

In this case, one can show that the transformations of the phase space functions are generated[11] by the gauge generator:

$$G = \int d^3x \left( \dot{\epsilon} C_1 - \epsilon C_2 \right). \quad (7)$$

and the arbitrary function $\lambda$ transforms as $\lambda \to \lambda + \dot{\epsilon}$. It is straightforward to see that this transformation is just the same gauge transformation of the original Lagrangian action and of the 4-momentum version of Maxwell equations, i.e., it transforms $A_\mu$ into $A_\mu + \partial_\mu \epsilon$.

In this sense, we see how the total Hamiltonian formalism preserves the same structures and symmetries that were present at the original 4-momentum and Lagrangian formalisms. The physical content of the theory, in any of its three formulations, is the electric and magnetic fields. In the total Hamiltonian formalism, the magnetic field is given by $\vec{\nabla} \times \vec{A}$, while for the electric field we have two variables playing that role, $\vec{\pi}$ and $\dot{\vec{A}} - \vec{\nabla} A_0$, in the same way that in the Hamiltonian formulation of the dynamics of a single particle both $p$ and $m\dot{q}$ represent linear momentum. These three phase space functions remain invariant under a transformation generated by the gauge generator and are thus considered observables. The other non-trivial phase space function which is invariant is the momentum $\pi^0$, which is constrained to be 0.

---

7    The implication does not work in both directions, as second-class systems will also have singular Lagrangians but no (genuine) gauge symmetry.

8    Rather than a constraint, it is an infinite number of constraints, one per space point, i.e., $C_1(\vec{x}) = 0$. However, many times one refers to this family of constraints simply as the primary constraint.

9    I refer the reader to Rothe and Rothe (2010) for a detailed derivation of the total Hamiltonian formalism.

10    The symbol $\approx$ means weak equality, i.e., equality on the constrained surface. In this case the equality is strong, as it holds in the whole phase space.

11    This means that the infinitesimal transformations are of the form $f(q,p) \to f(q,p) + \{f,G\}$. The finite version of these transformations is given by an "exponentiation" of the infinitesimal one. For the cases considered in this article, this distinction won't make a difference until I consider the general case in section 4.

For the case of other "genuine" gauge theories we find the same structure, i.e., a singular Lagrangian with a number $n$ of independent local symmetries, which correspond to $n$ primary constraints and which can be represented in the total Hamiltonian formalism using a canonical Hamiltonian plus these $n$ constraints multiplied by $n$ arbitrary functions. The original $n$ gauge symmetries correspond to the $n$ symmetries of the total action, which are generated by the $n$ generators. When there are secondary constraints, the form of these generators is that of a tuned sum of constraints[12] and they are not just simply any first-class constraint. First-class constraints, when secondary constraints are present, do **not** generate symmetry transformations of the total Hamiltonian nor symmetry transformations of the original Lagrangian action. To look for the physical content of the theory, one should look at what remains invariant under the transformations generated by the generators.[13]

In this sense, we can read Pitts (2014) as arguing precisely for this claim. To argue that every first-class constraint generates a gauge transformation, one needs to move to the extended formalism, which I will come back to discuss in section 3.2. For now, let me insist on the highlight of this discussion: if one has a gauge theory defined by means of a Lagrangian action showing just the "genuine" gauge symmetries of the theory, whichever way you define them, the proper way of preserving these symmetries in the Hamiltonian formalism is by means of the total Hamiltonian.

## 2.2 DE-OCKHAMIZATION

Brian Pitts suggests that, besides genuine gauge transformations, we find other transformations that can be argued to leave the physical content of the theories unaffected. However, these transformations lack any physical motivation and correspond to just an artificial overcomplication of our formalism. In this sense, one goes from a (relatively) simple physical theory to an unnecessarily complicated one, contrary to Ockham's razor, and hence the name. The well-discussed example he gives is the substitution of force by gorce plus morce, as introduced in (Glymour 1977) and discussed since.

Here I will take a simple example: the dynamics of a free particle in Newtonian physics. A free particle moves in a straight line at uniform velocity, as one can derive by minimizing the action:

$$S[q] = \int dt \, \frac{m}{2} \dot{q}^2. \tag{8}$$

Now someone comes along and defines a "gauge" transformation that adds to a given $q(t)$ an arbitrary function $\mu(t)$. That this is to be considered a gauge transformation strikes us as odd, as $q(t)$ goes from describing a trajectory with uniform velocity to any arbitrary function, no matter how wild. To this, the proponent of the transformation tells us that we are right, but that after the transformation we cannot interpret $q(t)$ as describing the trajectory of a free particle, and that we should take $q - \mu$ to represent the position of the new particle. Furthermore, they propose that we should also replace the original action with a new one:

$$S[q] = \int dt \, \frac{m}{2} (\dot{q} - \dot{\mu})^2. \tag{9}$$

---

12 Two different derivations of the relations that need to hold between the coefficients of the different constraints of the generator can be found in Rothe and Rothe (2010, Sects. 5.3 and 5.4).

13 This is not necessarily the same as saying that the observables of the theory are the phase space functions that remain invariant under a transformation, as the discussion of reparametrization invariant shows us that this is more subtle. I refer the reader to Mozota Frauca (2023), Pitts (2017; 2018, and Pons, Salisbury, and Sundermeyer (2010) for discussions of this point and for the distinction between global and instantaneous types of gauge transformations.

One can see that this new action has an explicit symmetry under the transformation $q \to q + \epsilon, \mu \to \mu + \epsilon$. The equations of motion for this Lagrangian are:

$$\frac{d}{dt}(\dot{q} - \dot{\mu}) = 0. \tag{10}$$

That is, $q - \mu$ describes particles moving at uniform velocities, as we were expecting. This, of course, works in order to keep the empirical content of the theory but, as Pitts argues, if this is to be considered a gauge transformation, it is only in a very trivial way, which does not have anything to do with any possible, preexisting genuine gauge symmetry.

In this sense, for any theory one can choose any arbitrary transformation and say it is a symmetry transformation or even a gauge transformation, as one only needs to introduce enough compensating functions to undo the transformation and to redefine or change the way the variables in the formalism correspond to physical quantities in the real world.

It is in this sense that Pitts emphasizes that, in his opinion, the claim that first-class constraints generate gauge transformations is either trivial or false. For the simple example I have introduced, this seems to be the case, as the transformation doesn't keep the physical content of the theory intact unless we reinterpret $q - \mu$ as describing the trajectory of the particle, and this is trivial in the sense that we have just introduced.

We can express this in the Hamiltonian language. The momentum conjugate to $q$ in the de-Ockhamized action 9 is:

$$p = \frac{\partial L}{\partial \dot{q}} = m(\dot{q} - \dot{\mu}). \tag{11}$$

This is just the physical linear momentum expressed in the de-Ockhamized variables. The Hamiltonian is:

$$H = \frac{p^2}{2m} + p\dot{\mu}. \tag{12}$$

Making use of the symplectic structure of the phase space, we can identify momentum $p$ as the "gauge" generator which generates the transformation $q \to q + \epsilon, p \to p$, as long as we accompany this with the transformation $\mu \to \mu + \epsilon$, just as in the case of "genuine" gauge transformations the gauge transformations needed of the combined action of the gauge generator and a change in the arbitrary functions accompanying the primary constraints, e.g., $\lambda$ in the case of electromagnetism.[14] The most striking difference with the case of 'genuine' gauge transformations is that the symmetry transformations do not need to be generated by constraints or combinations of constraints, any phase space function can generate a symmetry transformation if one just adds the pertinent compensating functions.

---

14     There is an alternative description of de-Ockhamizations in the language of phase spaces and constrained systems which consists of taking the compensating functions to be phase space variables. If one chooses that formalism, de-Ockhamizations correspond to transformations generated by $P_\mu + \phi$, where $P_\mu$ is the momentum conjugate to $\mu$ and $\phi$ the arbitrary phase space function that defines the de-Ockhamization. Despite this, I will stick to the formulation that takes compensating functions to be arbitrary functions and not phase space variables, as it is the formulation used in both Pitts (2022) and Pooley and Wallace (2022) for the discussion of electromagnetism.

It is in this sense that one can argue that any phase space function, including any constraint, generates a "gauge" transformation.[15] However, in general they won't correspond to any "genuine" gauge transformation, and one needs to change the physical interpretation one gives to the variables in the formalism and to change the equations of motion to include the effect of the compensating functions. All of this has an ad hoc feeling that justifies the claim that it is false or just trivially true that any phase space function generates a gauge transformation. That is, general phase space functions generate gauge transformations only in the trivial sense of gauge transformation.

## 2.3 MOMENTUM DE-OCKHAMIZATION

The example in this subsection aims to capture the intuition behind the "extended" sense of gauge transformation, despite not being related to an extended formalism of gauge theories and drawing intuition from different sources. However, it aims to illustrate how a de-Ockhamization can be considered not to fit in either the trivial or the genuine gauge transformation categories. The trick for this is that, while the transformation is formally a de-Ockhamization (and hence not a genuine gauge transformation), it is one that affects accessory variables and not physical variables. For this reason, one can ignore the effects of these transformations, contrary to what happened in the previous example, where we had to introduce a redefinition of our physical variables. It is in this sense that this kind of example is different from other de-Ockhamizations and can be considered to deserve a different category. In section 3.2, I will argue that the extended transformations of electromagnetism fit in this category.

In the example of a non-relativistic particle, let me adopt the following perspective, which is natural in the context of classical mechanics. We start with a theory, Newtonian mechanics, which states that free bodies move in straight lines with uniform velocities. In this sense, the physical content of the theory is given by the set of physically allowed trajectories $q(t)$, where $q$ represents the position of a particle or body in space. Now we can express this theory in the Hamiltonian formalism, where a variable $p$ will appear, but it can be consistently understood to be just an accessory variable: At the end of the day we just care about the trajectories $q(t)$, and $p$ is just a useful part of the mathematical machinery we use for computing them. From this point of view, $q$ is a physical variable while $p$ is an accessory variable.

If one accepts this perspective, then one may have different intuitions about de-Ockhamizations depending on which variables are affected by them. In the previous example, the de-Ockhamization affected $q(t)$, and we were forced to accept a redefinition in order to keep the physical content of the model. In the following example, $p$ is affected, but $q(t)$ is not, and hence one can claim that the physical content is not affected by this transformation and that this kind of transformation is different. Something similar will happen in the case of electromagnetism, as Pooley and Wallace will claim that the (according to them) physical content $\bar{\pi}$ is preserved even if the accessory $A_0$ is de-Ockhamized.

---

15    Let me make clear the difference between Pitts's position and mine: While Pitts argues that some constraints generate trivial gauge transformations, I am arguing for the more general claim that any phase space function generates a trivial gauge transformation, provided that the appropriate compensating functions are in place and that the new identifications and redefinitions are performed.

Having said this, let me introduce the example. Consider the following action for the Newtonian particle:

$$S[q] = \int dt \left[ \frac{m}{2} \dot{q}^2 + (\dot{\mu}q + \mu\dot{q}) \right], \tag{13}$$

where $\mu$ is an arbitrary function. From the Lagrangian perspective, not much has changed, as we have just added a total derivative term to the action 8. The equations of motion for $q$ are independent of $\mu$, and they describe particles moving uniformly in straight lines, as one could have expected. However, when we move to the Hamiltonian formalism we find that a de-Ockhamization has taken place, as momentum is now:

$$p = \frac{\partial L}{\partial \dot{q}} = m\dot{q} + \mu. \tag{14}$$

This is, we find that the momentum for this action has been shifted by an arbitrary function. The Hamiltonian becomes:

$$H = \frac{(p - \mu)^2}{2m} - q\dot{\mu}. \tag{15}$$

The equations of motion for this Hamiltonian are equivalent to the ones of the free particle in the sense that solutions for $q(t)$ still represent uniformly moving particles, although the equation for $p$ now has changed and depends on $\mu$. $p$ has lost its interpretation as the linear momentum, but this is a loss we may accept, as the physical meaning of canonical momentum variables, for standard, classical, non-gauge theories, is defined by means of the Lagrangian or, equivalently, of the Hamiltonian of the theory. That is, from the perspective we are taking in this example, momentum variables are accessory variables that draw their physical meaning from the Hamilton equation for velocities (which we assume to be physical).[16]

Indeed, if one works on the Hamilton equations of motion of this model, at the end of the day one is left with:

$$m\ddot{q} = 0, \tag{16}$$

that is, the equations of motion have not been affected by the de-Ockhamization and we find no trace of the compensating function $\mu$ in them, nor of the "accessory" momentum $p$. This is in contrast with the equations of motion of the model in the previous example 10, in which the compensating function appeared. This feature allows us to claim that the two de-Ockhamizations are fundamentally different: While one affects just accessory structures and leaves the physical quantities and equations of motion unaffected, the other one affects physical quantities and equations of motion and requires some redefinition in order to preserve the physical content.

This model has a symmetry transformation which consists of replacing $p$ with $p + \epsilon$ and $\mu$ with $\mu + \epsilon$. This transformation leaves the action invariant, and its phase space part is generated by $-q$. Calling this symmetry a "gauge" transformation goes against the conventional wisdom that there is no gauge symmetry in the dynamics of a free Newtonian particle, and it is certainly artificial in that we have introduced an arbitrary function. On the other hand, it is a transformation that leaves $q(t)$ (and its equations of motion) untouched, with no need to introduce a redefinition of the physical meaning of the configuration space variables of our model. There is a de-Ockhamization that only affects $p$, and hence it does not represent linear momentum anymore. However, as $m\dot{q}$ isn't affected, one can claim that the de-Ockhamization affects the accessory representation of linear momentum but not the linear momentum itself. In the case of electromagnetism, we will find a similar situation: We will find two expressions ($\vec{\pi}$ and $\dot{\vec{A}} - \vec{\nabla}A_0$), which represent the same physical quantity and the argument for considering that a de-Ockhamization is a gauge transformation will rely on the fact that only one of the two, which is considered accessory, is affected by the de-Ockhamization, while the other and the empirical content are arguably preserved.

---

16    For theories like electromagnetism this will be controversial, but I take it to be a reasonable position that for a theory describing the movement of bodies, the position of these bodies can be assumed to be physical and other variables acquire their meaning from their relations with them.

From this point of view, the claim that de-Ockhamizations represent bad physical changes or trivial transformations can be challenged in the case of momentum de-Ockhamization by adopting the position I have been taking in this section. As the transformation leaves $q(t)$ and its equations of motion intact, it is not a bad physical change. As one can ignore the effect of the transformation and one does not need to introduce compensating functions to extract the physical content of the theory, it is not a trivial transformation either. Therefore, one can hold that this kind of transformation lies in between completely trivial and genuine gauge transformation.

This kind of argument is the same that we will find in the case of electromagnetism for arguing that first-class constraints generate gauge transformations. To insist, the idea is that we have a family of transformations that are not of the original, Lagrangian, or "genuine" kind, as they carry with them a de-Ockhamization and the introduction of new variables. However, while for general de-Ockhamizations one needs to take care of these new variables and perform new identifications in order to keep the empirical content, in the case of electromagnetism and in the case of momentum de-Ockhamization this is not necessary. For this reason, one can argue that these transformations are part of a third category which is neither the original one, nor the category of trivial de-Ockhamizations. I will come back to this point when I discuss electromagnetism in the extended formalism in section 3.2.

# 3 TRANSFORMATIONS IN ELECTROMAGNETISM AND PROCA THEORY

Having introduced the three different kinds of transformations that one can consider to be gauge transformations, now we are in a position to analyze the two relevant cases for the controversy between Pitts and Pooley and Wallace in some detail. I will first analyze the case of Proca theory, and I will agree with Pitts in that the primary constraint of the theory generates a trivial de-Ockhamization. However, I will extend his claim and argue that the secondary constraint also generates a transformation that can be considered to be of the trivial kind of gauge transformation. In any case, I will agree with him in that the study of Proca's electromagnetism shows how a too-trivial notion of gauge transformation goes against the orthodox view, as the orthodox view claims that second-class constraints do not generate gauge transformations.

Then, I will move to the case of electromagnetism in the extended formalism and argue that one can see the transformations generated by the constraints in this case as pertaining to the third category of transformations. I will notice that for this, one needs to claim that momentum variables are physical while configuration variables are accessory (in this case), which implies an important departure from Lagrangian intuitions. Despite this, I will conclude that it is a consistent position and that it allows for saving the orthodoxy.

# 3.1 THE ACTION OF THE CONSTRAINTS IN PROCA THEORY

Proca theory is defined by the action:

$$S_{Proca}[A_\mu] = \int dt d^3x \left[ \frac{1}{2}(\dot{\vec{A}} - \vec{\nabla}A_0)^2 - \frac{1}{2}(\vec{\nabla}\times\vec{A})^2 - \frac{m^2}{2}A_\mu A^\mu - (A_0\rho + \vec{j}\cdot\vec{A}) \right]. \quad (17)$$

This Lagrangian is just the electromagnetic Lagrangian with the addition of the term $-\frac{m^2}{2}A_\mu A^\mu = \frac{m^2}{2}(A_0^2 - \vec{A}^2)$, where the parameter $m$ is the "photon mass." This term spoils the local symmetry of the Lagrangian, so there is no "genuine" gauge transformation now. If one derives the equations of motion for $A_\mu$, one can check that the electric and magnetic fields have not only charged matter as sources, but also that the very same $A_\mu$ plays this role. This means that, if we take two 4-potential configurations like $A_\mu$ and $A_\mu + \partial_\mu \epsilon$[17] and their velocities, which were physically equivalent in the Maxwell theory, as initial conditions at a time $t$, their time evolutions will differ, affecting also the electric and magnetic fields. In Proca theory, $A_\mu$ is a physical field and not a gauge field.

Nevertheless, Proca theory also has some properties of its close relative Maxwell electromagnetism. Its Lagrangian is also singular, it is also a theory with constraints, indeed with very similar constraints, and needs to be treated in the Hamiltonian formalism as a constrained system. Its total Hamiltonian density is:

$$H_T(\vec{A},\vec{\pi},A_0,\pi^0) = H_c(\vec{A},\vec{\pi},A_0) + \lambda\pi^0 = \frac{1}{2}(\vec{\pi}^2 + \vec{B}^2) - A_0(\vec{\nabla}\cdot\vec{\pi}) + \frac{m^2}{2}A_\mu A^\mu + (A_0\rho + \vec{j}\cdot\vec{A}) + \lambda\pi^0. \quad (18)$$

Its primary constraint is just the same as in Maxwell electromagnetism, $C_1 = \pi^0$, and imposing its constancy leads to the secondary constraint $C_2 = \vec{\nabla}\cdot\vec{\pi} - \rho + m^2 A_0$, which is the same constraint as in electromagnetism but with the addition of an extra term $m^2 A_0$.[18] This extra term makes the constraints second-class, as now we have $\{C_1(x), C_2(y)\} = -m^2\delta^3(x-y) \neq 0$. This implies that when we impose constancy of the secondary constraint what we obtain is the condition $\lambda = \vec{\nabla}\cdot\vec{A}$. This reflects the fact that Proca electromagnetism is not a gauge theory and that there is no room for arbitrary functions in this theory. On the same line, there is no local symmetry in the total action. In this sense, Proca theory is just like any non-gauge theory but with the added complication that it has constraints.

For this reason, it reflects the "conventional wisdom" that second-class constraints do not generate gauge transformations, as they appear in non-gauge theories. Pitts (2022) argues that, if one applies the reasoning of Pooley and Wallace (2022) to the case of Proca, one would have to claim that the constraint $C_1$ generates a gauge transformation. This is contrary to the "conventional wisdom" that Pooley and Wallace wanted to defend, and hence Pitts argues that, by trying to save the "first-class constraints generate gauge transformations" doctrine, they have ended up trivializing the notion of gauge and being forced to accept that second-class constraints would generate gauge transformations too. Now I will analyze in some detail this example and evaluate Pitts's claims. In particular, I will expand on his claim and I will argue that $C_2$ also generates a trivial gauge transformation. More importantly, I will also argue that Pooley and Wallace can resist Pitts's arguments and defend a notion of gauge transformation which does not entail that secondary constraints in Proca theory generate gauge transformations.

---

17    For both of them to be acceptable initial conditions to Proca equations, $\epsilon$ has to satisfy $\partial_\mu\partial^\mu\epsilon = 0$.

18    This is a version of Gauss law that takes $A_0$ to act as a source of the electric field.

Pitts shows that the action of $C_1 = \pi^0$ is to shift the 0th component of the 4-potential, $A_0$, and that, in Proca theory, this can be considered a symmetry transformation only if one introduces a de-Ockhamization and a compensating function $\mu_1$, such that it is $A_0 - \mu_1$ that ends up playing the role that $A_0$ was playing before. That is, the physically meaningful $A_0$ has been replaced by the now unphysical $A_0$ and $\mu_1$, just as in section 2.2 the physically meaningful $q$ was replaced by $q - \mu$. I completely agree with his analysis on this point: $C_1$ only generates gauge transformations of the trivial kind.

However, Pitts also claims that "only the primary (second-class) constraint, not the secondary (second-class) constraint, generates a gauge transformation (by the standards at hand)" (2022, 19), where he is referring to gauge transformations of the trivial kind. This claim is true if one is referring to the de-Ockhamized model he is discussing, but I will argue that it is false if one refers to other de-Ockhamized versions of Proca theory. In particular, the secondary constraint $C_2$ also generates a trivial gauge transformation, that is, it is a symmetry of a de-Ockhamized action in which a function $\mu_2$ is introduced to compensate the effect of the transformation. I will refer to this theory as the doubly de-Ockhamized Proca theory, as one needs to introduce two de-Ockhamizations. In this sense, what I am arguing for is what I anticipated in the introduction and in the discussion of the examples in section 2, that any phase space function can generate a trivial transformation if the right compensating functions are introduced appropriately. In section 4, I will give a more formal argument for the general claim.

Let me now give the expression of the doubly de-Ockhamized Lagrangian for Proca theory, which has as trivial gauge symmetries the symmetries generated by $C_1$ and $C_2$. This is an extension of the Lagrangian given by Pitts in (Pitts 2022) to include the de-Ockhamization associated with $C_2$. The doubly de-Ockhamized Lagrangian is:

$$L(A_0, \dot{A}_0, \vec{A}, \dot{\vec{A}}) = L_{Proca}(A_0 - \mu_1, \dot{A}_0 - \dot{\mu}_1, \vec{A} - \vec{\nabla}\mu_2, \dot{\vec{A}} - \vec{\nabla}\dot{\mu}_2) + m^2(\dot{\mu}_2 A_0 + \mu_2 \dot{A}_0), \quad (19)$$

where $L_{Proca}$ is the Lagrangian in the Proca action 17, and $\mu_1$ and $\mu_2$ are the compensating functions introduced to compensate the effects of the transformations generated by $C_1$ and $C_2$. There are two effects of the de-Ockhamization. First, when the de-Ockhamization affects configuration variables, the Lagrangian is modified to depend on the de-Ockhamized variables. That is, in the same way that in the example in section 2.2 the Lagrangian went from depending on $q$ to depending on $q - \mu$, in this case the Lagrangian depends on the de-Ockhamized version of the 4-potential $A_\mu$. Second, as the de-Ockhamization also affects the momentum variable $\pi^0$ (this will be clear below), we need to introduce a total derivative term to account for the change in momentum. This is just as in the example in section 2.3, where the term $\dot{\mu}q + \mu\dot{q}$ was introduced in order to de-Ockhamize $p$. In section 4, I will come back to comment on this general pattern.

By performing the Legendre transform to this Lagrangian one can find the Hamiltonian associated with this action:

$$H(\vec{A}, \vec{\pi}, A_0, \pi^0) = H_T(\vec{A} - \vec{\nabla}\mu_2, \vec{\pi}, A_0 - \mu_1, \pi^0 - m^2\mu_2) + \dot{\mu}_1 C_1 - \dot{\mu}_2 C_2 - \dot{\mu}_2(m^2\mu_1 + \rho), \quad (20)$$

where $H_T$ is the total Hamiltonian of the Proca theory (18)[19] but for the de-Ockhamized variables, that is, in this extended version $\vec{A} - \vec{\nabla}\mu_2$ plays the role of $\vec{A}$ in the original Proca theory, $A_0 - \mu_1$ plays the role of $A_0$, and $\pi^0 - m^2\mu_2$ plays the role of $\pi^0$ (and $\vec{\pi}$ stays the same). $C_1 = \pi^0 - m^2 \mu_2$ and $C_2 = \vec{\nabla} \cdot \vec{\pi} - \rho + m^2 (A_0 - \mu_1)$ are the constraints expressed in the de-Ockhamized variables. This is the Proca version of the extended Hamiltonian, which requires the introduction of two arbitrary functions (one per constraint), and requires reinterpreting the variables in the formalism according to the de-Ockhamization performed. The terms $-\dot{\mu}_2 (m^2 \mu_1 + \rho)$ do not play any role in the Hamiltonian dynamics.

---

19    As discussed above, $\lambda$ is not a free function but fixed to be $\vec{\nabla} \cdot \vec{A}$. In the de-Ockhamized version it becomes $\vec{\nabla} \cdot (\vec{A} - \vec{\nabla}\mu_2)$.

One can check that the dynamics defined by the Hamiltonian 20 or by the action built using the Lagrangian 19 is invariant under transformations generated by both $C_1$ and $C_2$ as long as $\mu_1$ and $\mu_2$ change accordingly. Explicitly, there is invariance under transformations of the form:

$$A_0 \rightarrow A_0 + \epsilon_1 \tag{21}$$

$$\pi^0 \rightarrow \pi^0 + m^2 \epsilon_2 \tag{22}$$

$$\vec{A} \rightarrow \vec{A} + \vec{\nabla} \epsilon_2 \tag{23}$$

$$\vec{\pi} \rightarrow \vec{\pi} \tag{24}$$

$$\mu_1 \rightarrow \mu_1 + \epsilon_1 \tag{25}$$

$$\mu_2 \rightarrow \mu_2 + \epsilon_2, \tag{26}$$

where $\epsilon_1$ and $\epsilon_2$ are arbitrary functions associated with the transformations generated by $C_1$ and $C_2$, respectively.[20]

In this sense, by going to this extended version of Proca theory, we have formulated a theory with two additional symmetries. Certainly, they are not genuine gauge transformations from the point of view of Pitts, as they are not there in the original Proca theory, which is not a gauge theory. But, now the relevant question is, are these transformations just trivial de-Ockhamizations, or are they something more interesting as the gauge transformations of electromagnetism, according to Pooley and Wallace? While Pitts argues that there is no difference between these artificial transformations for Proca theory and the extended transformations of electromagnetism, I will argue here and in the next subsection that there are important differences that allow us to consider them different kinds of transformations. That is, according to the "extended" notion of gauge transformations, one can consistently claim that the constraints in Proca electromagnetism do not generate gauge transformations, while the constraints in Maxwell electromagnetism do.

As discussed above, the key criterion for deciding when a transformation is a gauge transformation in the extended sense or a trivial de-Ockhamization is whether it is a transformation that affects the physical, empirical content of the theory or if it is a transformation that affects accessory structures. In the case of Proca electromagnetism, $A_\mu$ is not a gauge field but a physical field, as I have discussed above. As the transformations generated by $C_1$ and $C_2$ affect $A_\mu$, they cannot be said to affect just accessory variables, and one needs to introduce redefinitions and identifications in order to recover the physical content of the theory. Furthermore, this can be seen in that, in the equations of motion of this model, the fields $\mu_1$ and $\mu_2$ appear. For instance, the de-Ockhamized version of Proca's version of Gauss law is

$$\vec{\nabla} \cdot \vec{\pi} - \rho + m^2 (A_0 - \mu_1), \tag{27}$$

---

20    More explicitly, the form of the generator of the phase space part of this transformation is $G = \int d^3x \left( \epsilon_1(t,x) C_1(t,x) + \epsilon_2(t,x) C_2(t,x) \right)$.

and similar results can be found for Proca equation, the dynamic equation of the theory. In this sense, when we compare this case with the examples in sections 2.2 and 13 we find that this case is similar to the position de-Ockhamization $q \rightarrow q - \mu$, and hence that the transformations are gauge transformations only in the most trivial sense of the term.

However, I will show next that in the case of Maxwell electromagnetism, this won't be the case, which allows for a consistent way of keeping the orthodoxy.

## 3.2 ELECTROMAGNETISM IN THE EXTENDED FORMALISM

Now we can finally turn to the main case of disagreement between Pitts and Pooley and Wallace, electromagnetism expressed in the extended formalism. Both parties agree that the extended formalism in the case of electromagnetism is obtained by de-Ockhamizing the electromagnetic potential, i.e., by replacing $A_0$ with $A_0 - \mu$. The de-Ockhamized action[21] is thus:

$$S_{em}[A_\mu] = \int dt d^3x \left[ \frac{1}{2}(\dot{\vec{A}} - \vec{\nabla}(A_0 - \mu))^2 - \frac{1}{2}(\vec{\nabla} \times \vec{A})^2 - ((A_0 - \mu)\rho + \vec{j} \cdot \vec{A}) \right], \quad (28)$$

and the extended Hamiltonian density:

$$H_{em} = H_c(\vec{A}, \vec{\pi}, A_0) + \lambda' \pi^0 + \mu(\vec{\nabla} \cdot \vec{\pi} - \rho) = H_c(\vec{A}, \vec{\pi}, A_0 - \mu) + \lambda \pi^0 - \dot{\mu} \pi^0. \quad (29)$$

Here, the function $\lambda'$ is related to $\lambda$ in the original total action 6 by means of $\lambda' = \lambda + \dot{\mu}$ and it is also an arbitrary function reflecting the original gauge freedom of electromagnetism. Now we have enlarged the original symmetry group of the action and symmetry transformations are generated by the constraints $C_1$ and $C_2$ independently, as long as $\lambda'$ and $\mu$ change accordingly. In particular, the symmetry transformations are given by:

$$A_0 \rightarrow A_0 + \epsilon_1 \quad (30)$$

$$\pi^0 \rightarrow \pi^0 \quad (31)$$

$$\vec{A} \rightarrow \vec{A} + \vec{\nabla}\epsilon_2 \quad (32)$$

$$\vec{\pi} \rightarrow \vec{\pi} \quad (33)$$

$$\mu \rightarrow \mu + \epsilon_1 - \dot{\epsilon}_2 \quad (34)$$

$$\lambda' \rightarrow \lambda' + \dot{\epsilon}_1. \quad (35)$$

As before, $\epsilon_1$ and $\epsilon_2$ are arbitrary functions associated with the constraints $C_1$ and $C_2$ respectively. We can recover the original gauge symmetry by setting $\epsilon_1 = \dot{\epsilon}_2$.

In this de-Ockhamized, or extended, formalism we find the same physical quantities as in the original version of electromagnetism: We have the magnetic field as represented by $\vec{\nabla} \times \vec{A}$ and the electric field as represented by $\vec{\pi}$ and also by $\dot{\vec{A}} - \vec{\nabla}(A_0 - \mu)$. All these quantities are invariant under the transformations above and therefore "gauge" invariant. The second expression for the electric field carries a $\mu$-dependence and therefore is de-Ockhamization dependent, while the first one remains independent of that de-Ockhamization, as it is just $\vec{\pi}$. Moreover, one can express the Hamilton equations for $\vec{\pi}$ and $\vec{\nabla} \times \vec{A}$ just in terms of $\vec{\pi}$ and $\vec{\nabla} \times \vec{A}$ (and of $\vec{j}$), and this, together with the constraint $\vec{\nabla} \cdot \vec{\pi} - \rho = 0$ and the identity $\vec{\nabla} \cdot (\vec{\nabla} \times \vec{A}) = 0$, gives the four Maxwell equations with no need to change the way one interprets $\vec{\nabla} \times \vec{A}$ or $\vec{\pi}$.

---

21 While electromagnetism can be thought of as the $m \rightarrow 0$ limit of the Proca theory, we have seen that their symmetry structure is very different. For this reason, in order to have symmetry transformations generated by the two constraints in Proca theory we need to de-Ockhamize the model twice, while in the case of electromagnetism with just one de-Ockhamization we obtain the desired symmetry structure. For this reason, the model discussed in this subsection is not, in a straightforward sense, the $m \rightarrow 0$ limit of the model discussed in the previous section. It is possible to build a doubly de-Ockhamized version of electromagnetism that corresponds to that limit, but its discussion is conceptually very similar to what I discuss in this section. I am thankful to an anonymous reviewer for inviting me to consider this relationship between models.

This is clearly similar to the case of the momentum de-Ockhamization I discussed in section 2.3. For that example, we had that the linear momentum of the particle was represented by $m\dot{q}$ but also by the de-Ockhamized $p - \mu$. By choosing the non-de-Ockhamized version, we didn't need to worry about how de-Ockhamization affected the physical meaning of $p$, and indeed at the end of the day the equations of motion for the physical $q$ were unchanged. This was straightforward in the Lagrangian formulation and required using all the equations of motion, including the ones for $p$, in the Hamiltonian formulation. Based on the primacy of the Lagrangian formalism and on our interpretation of the theory, $q$ was considered a physical variable and $p$ and $\mu$, just accessory variables.

In the case of the extended formalism for electromagnetism, one could make a similar claim: that $\vec{\pi}$ and $\vec{\nabla} \times \vec{A}$ (together with $\rho$ and $\vec{j}$) are the physical variables in the formalism while $A_0, \pi^0$, and the curl-free component[22] of $\vec{A}, \lambda'$ and $\mu$ are just accessory variables. This can be motivated by the observation that $\vec{\pi}$ and $\vec{\nabla} \times \vec{A}$ do, indeed, behave like the electric and magnetic fields, but it is relevant to note that now we are inverting the role of momentum and configuration variables: While in the particle case we were claiming that the configuration space variables were the physical ones and the momentum variables were accessory variables, now it is (some) configuration variables that we would claim that are "accessory."

Pooley and Wallace adopt a perspective in which $\vec{\pi}$ is considered physical and that the transformations generated by the constraints should be considered gauge transformations, despite the fact that there is a de-Ockhamization and an expansion of the formalism. At the same time, I think it is not necessary for them to embrace the most trivial view of gauge transformations, as Pitts's (2022) arguments were pushing them to embrace. There is a definition of gauge transformation, the extended view, as I presented it in the introduction, which is consistent and allows the saving of the orthodoxy. From the extended perspective, when the de-Ockhamization or extension does not affect the physical variables and their equations of motion, one can consider that the symmetry transformations associated with it are gauge transformations in this extended sense. In the case of electromagnetism, as de-Ockhamization leaves $\vec{\pi}$, $\vec{\nabla} \times \vec{A}$, and their equations of motion unaffected, one can adopt the extended view of gauge transformations by considering that $\vec{\pi}$ and $\vec{\nabla} \times \vec{A}$ represent the physical content of electromagnetism.

Pitts argued against this possibility (2014), as it represents an important departure from the Lagrangian understanding of classical theories. I refer the reader to Curiel (2014) for a discussion of classical systems and for an argument for why they are Lagrangian rather than Hamiltonian, but it is the case that, for generic classical theories, momentum variables and phase space are defined starting from a Lagrangian. In this sense, the most natural reading of classical theories is from a Lagrangian point of view, and the Hamiltonian formulation and variables are just convenient rewritings of the Lagrangian dynamics. From this perspective, claiming that some momentum variables are physical while some configuration space variables aren't, seems odd, and this is what Pitts argued in 2014.

---

22    That is, the part that does not contribute to $\vec{\nabla} \times \vec{A}$ and which is not invariant under the transformations above.

However, we are not dealing with a generic classical theory but with a gauge theory. In the 4-potential formulation of electromagnetism, one can challenge the claim that configuration space variables have physical meaning, as it is $\vec{\nabla} \times \vec{A}$ and $\dot{\vec{A}} - \vec{\nabla} A_0$ which do, and the latter lives in the tangent bundle and not in configuration space. In this sense, it is true that, in a sense, $A_\mu$ are accessory variables used to encode the physical $\vec{\nabla} \times \vec{A}$ and $\dot{\vec{A}} - \vec{\nabla} A_0$, and when one moves to phase space this is even stronger as I have commented above that the evolution of $\vec{\pi}$ and $\vec{\nabla} \times \vec{A}$ is independent of $A_0$, the curl-free component of $\vec{A}, \lambda'$ and $\mu$. What is more questionable is the claim that the physical meaning of $\vec{\pi}$ is independent of $A_\mu$ if we want to have the full analogy with the momentum de-Ockhamization, where the meaning of $q$ is independent of $p$.

One way to go would be just to postulate that $\vec{\pi}$ is the electric field, but Pitts rejects this on the grounds that "the electric field is what pushes on charge." However, Pooley and Wallace argue that, on the extended formalism, once the Lagrangian or Hamiltonian for matter is added, $\vec{\pi}$ is what appears in the equations of motion of matter playing the role of the electric field and, hence, that it is "what pushes on charge." This is not exactly true, as what appears playing the role of the electric field is $\dot{\vec{A}} - \vec{\nabla}(A_0 - \mu)$ as can be seen by deriving the Lorentz force expression for a charged particle using the (de-Ockhamized) action:

$$S[\vec{x}] = \int dt \left( \frac{1}{2} m\dot{\vec{x}}^2 - q(A_0(\vec{x}) - \mu(\vec{x}) + \dot{\vec{x}} \cdot \vec{A}(\vec{x})) \right). \tag{36}$$

This result is obtained independently of whether one directly uses this Lagrangian or instead one finds the Hamiltonian and then uses Hamilton equations. Similarly, Pooley and Wallace find that the same holds for an example using a matter field. The same is expected to happen for any generic matter theory: If one starts with an action depending on $A_\mu$ and which leads to equations of motion only depending on the electric field $\dot{\vec{A}} - \vec{\nabla} A_0$ and magnetic field $\vec{\nabla} \times \vec{A}$ and one de-Ockhamizes the potential $A_0$, then one will find that the de-Ockhamized electric field is what pushes charges around.

However, it is not immediate that one can equate the de-Ockhamized electric field with $\vec{\pi}$. For doing so, we need to use one of the Hamilton equations, and therefore one can argue that the meaning of $\vec{\pi}$ depends on $A_\mu$ after all, or at least on the Hamiltonian. This is clearly different from the case of $q$ in the case of a single particle, as one can wildly change the dynamics of that theory, i.e., its laws of motion, its Lagrangian or its Hamiltonian, that one can still interpret $q$ as describing the trajectory of a single particle. If we consider instead a different dynamics for the electromagnetic field but the same coupling to matter, we will find that it is still $\dot{\vec{A}} - \vec{\nabla}(A_0 - \mu)$ that plays the role of the electric field in pushing charged matter, even if its dynamics may not obey Maxwell equations anymore. Meanwhile, $\vec{\pi}$ will cease to be equal to the electric field.

In this sense, one can conclude that, while configuration space variables for a classical theory have physical meaning which is quite theory independent, the meaning of $\vec{\pi}$ is not, and would only ascribe physical meaning to it by means of the theories where it appears, just as in the case of any other momentum variable. This is, of course, valid as long as the way we couple matter to the electromagnetic field in the Lagrangians and Hamiltonians we use via the configuration variables $A_\mu$.

This sort of worry shows that there is a difference between the de-Ockhamized electromagnetism and the example of momentum de-Ockhamization and adds up to the general Lagrangian worry of Pitts. Despite this, the "extended" view of gauge transformations remains a consistent view of gauge transformations which, no matter how natural we find it, is able to give a definition of gauge transformation that does not fall into the trivial category. Moreover, even if from a Lagrangian perspective it seems that there is no reason for adopting it, I will argue in section 5 that a reason for adopting it lies in the quantization of gauge theories. Before this, in the next section I study how the analysis extends from the examples considered here to the general case.

# 4 CONSEQUENCES FOR THE DIRAC'S CONJECTURE AND THE EXTENDED FORMALISM

The discussion above should have made clear the sense in which one can claim that all first-class constraints generate gauge transformations in the case of electromagnetism while still claiming that second-class constraints or arbitrary phase-space functions don't. However, Dirac's conjecture wasn't concerned just with electromagnetism, but with any generic first-class gauge theory, in particular with those with secondary constraints, which are the controversial ones. What can we say about the general case in light of the above discussion?

In this section, I will argue for three general claims. First, for any theory and phase space function, we can define a de-Ockhamization which leads to a theory with symmetry transformations generated by that phase space function. Second, when the theory is a gauge theory, this procedure leads to Hamiltonians of the extended form, at least to first order. And third, Dirac's conjecture seems plausible when some restrictions are in play. That is, it seems that for any generic first-class system the first-class constraints generate gauge transformations in the extended sense. This means that, even if they are associated with a de-Ockhamization, it seems plausible that physical variables and their equations of motion aren't affected by this de-Ockhamization. For this, it will be generally the case that some phase space momenta will be considered physical and that some configuration variables will be considered accessory, just as in the case of electromagnetism. In any case, note that my arguments for the case of Dirac's conjecture are just some plausibility arguments and not rigorous proof.

In this article, I have claimed several times that any phase space function can generate a gauge transformation if the necessary compensating functions are introduced. Now, I will give a construction of how this can be achieved for a generic theory. I will call $Q, P$ the phase space functions prior to the de-Ockhamization, and $q, p$ the phase space functions after a de-Ockhamization generated by a phase space function $\phi$.[23] The relation between these variables is:

$$Q = (\exp - \mu\{\cdot, \phi\})q = q - \mu\{q, \phi\} + \frac{\mu^2}{2}\{\{q, \phi\}, \phi\} + O(\mu^3) \tag{37}$$

$$P = (\exp - \mu\{\cdot, \phi\})p = q - \mu\{p, \phi\} + \frac{\mu^2}{2}\{\{p, \phi\}, \phi\} + O(\mu^3), \tag{38}$$

where the exponential of the Poisson bracket, $\exp - \mu\{\cdot, \phi\}$, is defined as a series in which the $n$-th term implies taking the Poisson bracket with $\phi$ $n$ times and $\mu$ is an arbitrary function of time or spacetime. This definition is analogous to the way the exponential of an operator is defined in quantum mechanics. This expression makes it clear that in the examples in this article (in both sections 2 and 3), it was enough to keep just the first order, given that, for simple constraints like the ones considered terms like $\{\{q, \phi\}, \phi\}$, containing multiple Poisson brackets would vanish.

---

23   The presentation in this section is in terms of generic Hamiltonian systems, the generalization to field theory is straightforward, and involves a spatial smearing of the constraints and functions generating de-Ockhamizations and gauge transformations, just as in the discussion in the previous section.

This definition is such that the result is invariant under a transformation generated by $\phi$, and which involves the appropriate change in $\mu$, as it is easy to check. That is, under a transformation:

$$q \rightarrow (\exp \epsilon \{\cdot, \phi\}) q \tag{39}$$

$$p \rightarrow (\exp \epsilon \{\cdot, \phi\}) p \tag{40}$$

$$\mu \rightarrow \mu + \epsilon, \tag{41}$$

$Q$ and $P$ do not change. This is precisely the symmetry transformation found in the examples in this article.

Now let me show the way in which the Hamiltonian action for $Q, P$ changes when one introduces the variables $q, p, \mu$. We start with an action principle of the form:

$$S[Q, P] = \int dt \left( P\dot{Q} - H(Q, P) \right). \tag{42}$$

And now, we simply replace $Q, P$ by their de-Ockhamized expressions $(\exp -\mu\{\cdot, \phi\}) q$, $(\exp -\mu\{\cdot, \phi\}) p$. By working on the term $P\dot{Q}$ one can express it in the following way,[24] up to a total derivative term:

$$P\dot{Q} = p\dot{q} - \dot{\mu}\phi - \mu \partial_t \phi + O(\mu^2). \tag{43}$$

This allows rearranging terms in the action so that it takes a Hamiltonian form:

$$S[q, p] = \int dt \left( P\dot{Q} - H(Q, P) \right) = \int dt \left( p\dot{q} - H_{ext}(q, p, \mu) \right), \tag{44}$$

where the "extended" Hamiltonian is:

$$H_{ext}(q, p, \mu) = H((\exp -\mu\{\cdot, \phi\}) q, (\exp -\mu\{\cdot, \phi\}) p) + \dot{\mu}\phi + \mu \partial_t \phi + O(\mu^2) \tag{45}$$

This is the form of the "extended" Hamiltonians in sections 2 and 3.1, up to a total derivative term in the last case and taking into account that the simple form of the constraints makes it the case that no higher order term in $\mu$ appears. If now one finds $p$ such that it minimizes the action and substitutes it in the action, one finds the Lagrangian expression of the de-Ockhamized theory. This shows what I have claimed before; one can de-Ockhamize any theory using any phase space function to generate such a de-Ockhamization: One just needs to replace the original variables $Q, P$ with their de-Ockhamized expressions, which will generally depend on the compensating function introduced $\mu$.

Now we can turn to the second general claim I want to argue for in this section. We are interested in the case in which $\phi$ is a first-class constraint and not just an arbitrary phase space function. That is, we are interested in de-Ockhamizing using the constraints, so that they will generate symmetry transformations.

The total Hamiltonian of a generic first-class system is:

$$H(Q, P) = H_c(Q, P) + \lambda_A \phi_1^A, \tag{46}$$

---

24   For this result, I am building on expression 5.16 in Rothe and Rothe (2010, 73), but allowing for the constraints to include an explicit time dependence, which explains the $\mu \partial_t \phi$ term that doesn't appear on that expression.

where $\phi_1^A$ represent the primary constraints of the system and I am using the convention that a repeated index represents a summation. We will be allowing for the presence of secondary constraints that arise when imposing that the constraints do not evolve in time and, making use that the system is first class, i.e., that the Poisson bracket of any two constraints weakly vanishes.[25] Secondary[26] constraints can be thus defined as

$$\phi_n^A = \{\phi_{n-1}^A, H_c\} + \frac{\partial \phi_{n-1}^A}{\partial t}. \tag{47}$$

Having introduced this, let me de-Ockhamize this system using as a generator the constraint $\phi_1^1$. Applying expression 45 we find that the extended Hamiltonian is:

$$H_{ext}(q,p,\mu) = H_c(Q,P) + \lambda_A \phi_1^A(Q,P) + \dot{\mu}\phi_1^1 + \mu \partial_t \phi_1^1 + O(\mu^2) \tag{48}$$

To obtain an expression similar to the extended Hamiltonian of electromagnetism (29) we need to expand the total Hamiltonian, i.e., the canonical Hamiltonian and the constraints, in powers of $\mu$. To first order, this gives:

$$H_c(Q,P) + \lambda_A \phi_1^A(Q,P) = H_c(q,p) + \lambda_A \phi_1^A(q,p) - \mu\{H_c, \phi_1^1\} - \lambda_A \mu\{\phi_1^A, \phi_1^1\} + O(\mu^2) \tag{49}$$

The term $-\mu\{H_c, \phi_1^1\}$ combines with the term $\mu \partial_t \phi_1^1$ to give rise to a term containing the secondary constraint as defined by expression 47. This leads to the extended Hamiltonian:

$$H_{ext}(q,p,\mu) = H_c(q,p) + \lambda_A \phi_1^A(q,p) + \dot{\mu}\phi_1^1 + \mu\phi_2^A + \lambda_A \mu\{\phi_1^1, \phi_1^A\} + O(\mu^2). \tag{50}$$

Note that this is exactly the form of the extended Hamiltonian of electromagnetism 29, as there was only one primary constraint and there were no tertiary or higher-order constraints.

Now, one could iterate and de-Ockhamize using some other primary constraint (if there were) such as $\phi_1^2$, or using a secondary constraint, although this de-Ockhamization may be uninteresting if there are no higher-order constraints. For instance, in the case of electromagnetism, once one has de-Ockhamized using the primary constraint $C_1$, further de-Ockhamizing using $C_2$ just leads to the same extended Hamiltonian but with a redefinition of $\lambda'$ and $\mu$, i.e., one has still two "free" functions, one of which can be used for recovering the variables in the original total formalism. Once every possible de-Ockhamization has been performed, it seems that one would end up with an extended Hamiltonian of the form:

$$H_{ext}(q,p,\mu) = H_c(q,p) + \lambda_A^n \phi_n^A(q,p) + O(\mu_i^2), \tag{51}$$

where the indices of $\lambda_A^n$ now include all secondary constraints and the $\lambda_A^n$ are now functions of the original free $\lambda_A$, linear in the de-Ockhamizing free functions $\mu_i$ (one per secondary constraint), and possibly also functions of the phase space coordinates $q,p$. One can expect the higher-order terms to be also some function of the constraints and, hence, the final form of the extended Hamiltonian is precisely the extended Hamiltonian of Dirac's conjecture:

$$H_{ext}(q,p,\mu) = H_c(q,p) + \lambda_A^n \phi_n^A(q,p), \tag{52}$$

---

25    That is, $\{\phi_n^A, \phi_m^B\} = K_{nm,C}^{AB} \phi_p^C$, where $K$ are some functions.

26    Here I am using "secondary" to refer also to tertiary, quaternary, and any $n$-ary constraints of the system.

where the $\lambda$ now can be more complicated functions of the $\mu_i$. As I said above, this isn't any formal proof but just a plausibility argument. However, note that for cases with simple, linear constraints such as the ones in the examples considered one doesn't need to worry about higher-order terms and it is true that de-Ockhamization of first-class systems leads to an extended Hamiltonian. In this sense, I consider that my second claim is plausible, i.e., that extended Hamiltonians of first-class systems are associated, at least to first order, with de-Ockhamizations generated by the constraints, which are symmetry generators of the de-Ockhamized system.

If the above is true, then for any extended system one should be able to express the original $Q, P$ in terms of $q, p$, and $\lambda_A^n$, and they should be invariant under transformations generated by the constraints when the $\lambda_A^n$ transform appropriately. As discussed in the previous section, this is enough to satisfy the most trivial definition of gauge symmetry and not a gauge symmetry from the most strict point of view. For the orthodox view to be true, as I have discussed in the previous section, what one needs is that these symmetry transformations are not just trivial transformations, but also "extended" gauge transformations. For this, I have argued that the physical variables (which one may need to argue they may be momentum and not configuration variables) and their equations of motion have to be independent of the de-Ockhamization. My third claim in this section is that, once some restrictions are in place, it seems plausible that this is the case and hence that Dirac's conjecture is true from the extended perspective.

Let me start the analysis of this issue by considering a first-class system with a primary, a secondary, and a tertiary constraint, and by assuming that the gauge symmetry of this system can be understood from a local point of view. That is, this system is not a reparametrization invariant model like general relativity and it makes sense to speak about what is observable at a time for this system.[27] To simplify, let's assume that all of the constraints have strongly vanishing Poisson brackets among themselves. In this case, the symmetry generator of the total Hamiltonian[28] would be:

$$G = \ddot{\epsilon} C_1 + \dot{\epsilon} C_2 + \epsilon C_3. \tag{53}$$

Invariant quantities have vanishing brackets with each of the constraints or involve some time derivatives, just as was the case for $\vec{A} - \vec{\nabla} A_0$ in the case of electromagnetism. But, now we can even have second temporal derivatives, which complicates the case. In the case of electromagnetism we could express $\vec{A} - \vec{\nabla} A_0$ just as $\vec{\pi}$, but when terms involving two temporal derivatives of configuration variables, or one temporal derivative of momentum variables, are present, invariant quantities cannot be expressed just as pure phase space functions, i.e., they will necessarily involve some time derivative. In this case, if we assume that the invariant quantities involving second temporal derivatives are part of the physical content of the original theory, then they are left out if someone claims that it is just quantities with vanishing Poisson brackets with the constraints that capture the physical content of the theory.

This shows that some limitation needs to be imposed. In the case in which the gauge invariant content at a time (or spacetime point) of the original Lagrangian formulation can be encoded just as a combination of configuration space variables and velocities, then they can be expressed as phase space functions with no temporal derivative involved and, as they must be invariant under the action of the generator, they will then have vanishing Poisson brackets with all the constraints separately. It is just in this kind of theory that the definition of observable as the function that has vanishing Poisson brackets with all the first-class constraints can apply. In this case, one could argue on the same lines as I have discussed in the previous section that these functions are the physical observables[29] of the theory and that they are preserved in the extended formalism, even if their counterparts, expressed as functions of configuration space variables and velocities, are de-Ockhamized. As in the case of electromagnetism, this entails a departure from the Lagrangian formalism as Pitts noted, but it is a consistent definition of gauge transformation that would allow to preserve the orthodoxy.

---

27    Again, see the discussion in Mozota Frauca (2023) and Pitts (2017; 2018).

28    An example of a Lagrangian with this constraint structure can be found in Castellani (1982, Sect. 4). This Lagrangian does not correspond to any physically interesting model.

Finally, we can also briefly comment on the case of reparametrization invariant theories like general relativity.[30] In this case, as I have mentioned in the introduction, it has been argued that the notions of gauge transformation at a time and observable may not make sense. However, from the point of view that takes transformations to be transformations between solutions of the equations of motion it may be the case that an extended sense of gauge transformation is available. For instance, in the case of general relativity each gauge generator is a combination of two types of constraints, one that affects the geometrical tensors ($g_{ab}, K_{ab}$) and another one that affects the lapse function and shift vector ($N, \vec{N}$). It is plausible to say that the primary constraint acts as a de-Ockhamization of this latter set of functions and that the physical content of the theory can be read just from the geometrical ones.[31] In this sense, it could be the case that Dirac's conjecture is right even in the case of general relativity, in that first-class constraints generate gauge transformations, in the global (as opposed to spatiotemporally local) and extended senses.[32] This is perfectly compatible with the claim that in the context of general relativity it does not make sense to define "observables" to be gauge invariant quantities at a spacetime point.

For these reasons, I believe that Dirac's conjecture is possibly true for the kind of theories I have discussed here[33] and once we adopt the extended notion of gauge. As I said above, my arguments give support to this possibility but they do not provide strict proof.

## 5 QUANTIZATION

Finally, in this last section I want to come back to the original motivation to introduce the extended Hamiltonian formalism. As I have argued above, moving to the extended formalism is probably equivalent to performing a de-Ockhamization, although rigorous proof for this seems to be missing. From the point of view of the classical theory there do not seem to be many reasons for adopting this formalism, and maintaining the Lagrangian-Hamiltonian equivalence would be a strong reason for rejecting it. One reason for adopting the extended formalism would be a preference for expressing gauge-invariant quantities just as phase space functions and not as temporal derivatives of these functions. Now, when we move to canonically quantizing the theory, one could argue that this preference is indeed an obligation, as quantum observables in the quantum theory correspond to phase space functions in the classical one. Moreover, the way constraints are imposed in the quantum formalism naturally leads to the extended notion of gauge transformation. In this section I will expand on these claims, showing that the way the canonical quantization of gauge systems is performed is most naturally understood from the extended view of gauge.

---

29    Note, however, that my argument starts from an analysis of the classical theory and its physical content (maybe even before expressing it in a Lagrangian formulation) and then studies in which conditions phase space function capture this content. This is opposite to the analysis one sometimes finds in the literature, in which it is claimed that the formal properties of phase space functions determine whether they are observable or not.

30    A formal discussion of this kind of system in the constrained Hamiltonian formalism can be found in Pons et al. 1997.

31    More precisely, given the relations between $g_{ab}, K_{ab}$ and $N, \vec{N}$, it seems likely that one can recover $N(t,x)$ and $\vec{N}(t,x)$ from $g_{ab}(t,x)$ and $K_{ab}(t,x)$. This is related to the thin sandwich conjecture.

32    During the review process of this article, Pitts has published a new article (Pitts 2024) in which he analyzes the case of massive theories of gravity. These theories stand in the same relation to general relativity as Proca theory to Maxwell's electromagnetism. For this reason, the analysis in this case is analogous. That is, Pitts argues that in the massive case a second-class constraint can be argued to generate a gauge transformation of the trivial kind. As I have argued for the case of electromagnetism, one can find important differences between the massive and massless cases, which allow defending that there is an extended sense of gauge transformation that applies to the transformations defined in the massless case and not in the massive models.

33    There were some possible counterexamples to Dirac's conjecture (Henneaux and Teitelboim 1992) that were argued not to be counterexamples in Rothe and Rothe (2010). It is beyond the scope of this article to discuss these examples, but the fact is that they are rather artificial, unphysical, and different from standard gauge theories.

In Dirac's quantization procedure for gauge theories (1964), constraints are imposed by requiring that physical states satisfy

$$\hat{\phi}_\alpha \psi(q) = 0 \, \forall \alpha, \tag{54}$$

where $\hat{\phi}_\alpha$ are the operator counterparts of the classical constraints, both primary and secondary, and $\psi(q)$ is a wavefunction(al) defined on the configuration space of the original theory. States satisfying the constraint equation are invariant under transformations of the form:

$$\psi \to e^{\epsilon^i \hat{\phi}_\alpha} \psi. \tag{55}$$

It is in this sense that it is very natural to consider that the quantum counterparts of the first-class constraints generate gauge transformations and that physical states are just gauge-invariant states. In this sense, this way of quantizing gauge theories very naturally fits with the extended definition of gauge in the classical theory.

As I have argued above, the extended formalism is related to a de-Ockhamization, and therefore it is interesting to study how de-Ockhamized theories could be quantized in analogous ways to Dirac's quantization. Let me consider the quantization of the single particle example above. Before de-Ockhamization the quantum theory describes the evolution of states $\psi(q)$ under the action of the quantum counterpart of the Hamiltonian. When we introduce a position de-Ockhamization we introduce a "gauge" symmetry generated by $p$ in the classical theory. It would be absurd to impose invariance under a transformation of the form $\psi(q) \to e^{\epsilon \hat{p}} \psi(q)$ as it leads to $q$-independent states. Instead, one would have to recall that after de-Ockhamizing, $q$ has lost its original meaning, and treat it in some appropriate way. For instance, one can fix $\mu$ and define states of the form $\psi_\mu(q)$. The physical position operator would now correspond to $\hat{q}_\mu = \hat{q} - \mu$. Under a "gauge" transformation we would have that states transform in the following way:

$$\psi_\mu(q) \to e^{\epsilon \hat{p}} \psi_\mu(q) = \psi_\mu(q + \epsilon) = \psi_{\mu+\epsilon}(q). \tag{56}$$

The initial and final states are different states, but the operators $\hat{q}_\mu$ and $\hat{q}_{\mu+\epsilon}$ have identical expectation values for the states $\psi_\mu(q)$ and $\psi_{\mu+\epsilon}(q)$, respectively.[34] This shows that, for a de-Ockhamization, imposing invariance under the generating operator is too strong a requirement, as one generally needs to consider the presence of the compensating functions and the way they affect the interpretation of the variables in the formalism.

---

34    There is an alternative version of this argument in a Heisenberg-picture-like version of gauge transformations, i.e., in a version in which it is operators that are affected by transformations and not states. While $\hat{q}$ is not invariant under a transformation generated by $\hat{p}, \hat{q}_\mu$, it is invariant once we take into account that $\mu$ also transforms to $\mu + \epsilon$.

In the case of electromagnetism, we could try to apply this lesson. This means that as the constraints $C_1$ and $C_2$ are related with de-Ockhamizations of $A_0$, one shouldn't impose invariance under the quantum transformations they generate. However, given that they are constraints, if we follow Dirac's quantization procedure we are forced to impose such an invariance. Luckily, even if in the particle case imposing invariance under the operator generating the de-Ockhamization leads to disaster, in the case of electromagnetism this is not the case. The reason for this is that the physical content of electromagnetism is captured by variables that are not affected by the de-Ockhamization. In the quantum case this means that there exist operators associated with $\vec{\pi}$ and $\vec{\nabla} \times \vec{A}$, and their quantum dynamics give rise to a meaningful quantum theory.

In this sense, we see how the quantization of electromagnetism using Dirac's extended formalism supports the extended view of gauge and distinguishes it from general de-Ockhamizations. While Pitts was right that the transformations generated by $C_1$ and $C_2$ in the classical case are associated with artificial de-Ockhamizations, in the quantum case they are part of the machinery that is used for defining the theory in a way that trivial de-Ockhamizations are not. For this reason, we can see how the extended view of gauge transformations makes the most sense in light of how quantum theories are customarily built.

In the case of more general theories, it seems that the same conclusion will obtain. For classical theories in which we can express the physical content of the theory as a set of phase space variables invariant under the extended set of gauge transformations, then Dirac's quantization procedure will preserve them, and from the quantum perspective the extended sense of gauge transformation will be natural. For reparametrization invariant theories, Dirac's quantization procedure is problematic for other reasons,[35] and hence the extended sense of gauge transformation cannot be supported by the quantum theory.

This section has shown how, even if from a classical point of view, the "genuine" sense of gauges is more natural, when we move to the quantum version of gauge theories the "extended" view of gauge transformations appears to fit very naturally with the way these quantum theories are defined.

## 6 CONCLUSIONS

In this article I have built on the controversy between Pitts and Pooley and Wallace to distinguish between three possible notions of gauge transformation. Pitts argued that there exist "genuine" gauge transformations, which are usually represented by symmetries of minimal, not artificially expanded Lagrangian actions, and then symmetries of artificial de-Ockhamizations or expansions of any theory. Given this classification, Pitts argued that Pooley and Wallace were forced to accept that the constraints in electromagnetism can only be said to be gauge transformations in the trivial sense, and that this further implied that the orthodox view couldn't be saved, as accepting the trivial sense of gauge transformation would entail that second-class constraints also generate gauge transformations. However, I have argued that there is room for a third sense of gauge transformation that allows escaping Pitts's dilemma.

This sense of gauge transformation is the "extended sense." This notion of gauge transformation accepts a departure from the original action and the introduction of compensating functions. However, these functions do not affect the physical variables, contrary to what happens in the case of a de-Ockhamization. To argue that this is the case, one needs to adopt a position in which momentum variables may be considered physical and configuration variables are considered accessory, which goes against many shared intuitions about classical theories. However, this is a consistent position that allows maintaining the orthodoxy. Furthermore, this extended view of gauge transformations fits nicely with the way gauge transformations are defined in quantum theories, and from that point of view it may not be such an unnatural definition.

---

35    I have argued for this claim in Mozota Frauca (2023).

Finally, I have also made some general claims, such as that any phase space function in any theory can be said to generate a trivial gauge transformation or that Dirac's conjecture may be true for generic gauge theories in the extended sense.

## ACKNOWLEDGEMENTS

I want to thank Brian Pitts and Carl Hoefer for their insightful comments and discussions.

## COMPETING INTERESTS

## AUTHOR AFFILIATIONS

**Álvaro Mozota Frauca**  orcid.org/0000-0002-7715-0563
Department of Physics, Universitat de Girona, Carrer de la Universitat de Girona 1, 17003 Girona, Spain